# Deep Learning for EELS hyperspectral images unmixing - using autoencoders

N. Brun*[1], G. Lambert[1] and L. Bocher[1]

1. Université Paris-Saclay, CNRS, Laboratoire de Physique des Solides

91405, Orsay, France

*corresponding author
nathalie.brun@universite-paris-saclay.fr
https://orcid.org/0000-0001-5519-3063

**Abstract**

Spatially resolved Electron Energy-Loss Spectroscopy (EELS) conducted in a Scanning Transmission Electron Microscope (STEM) enables the acquisition of hyperspectral images (HSIs). Spectral unmixing (SU) is the process of decomposing each spectrum of an HSI into a combination of representative spectra (endmembers) corresponding to compounds present in the sample along with their local proportions (abundances). SU is a complex task, and various methods have been developed in different communities using HSIs. However, none of these methods fully satisfy the STEM-EELS requirements. Recent advancements in remote sensing, which focus on Deep Learning techniques, have the potential to meet these requirements, particularly Autoencoders (AEs). In this study, the performance of Deep Learning methods using AE for SU is evaluated, and their results are compared with traditional methods. Synthetic HSIs have been created to quantitatively assess the outcomes of the unmixing process using specific metrics. The methods are subsequently applied to a series of experimental data. The findings demonstrate the promising potential of AE as a tool for STEM-EELS SU, marking a starting point for exploring more sophisticated Neural Networks.

## 1. Introduction

Continuous improvements in Scanning Transmission Electron Microscopes (STEM) and Electron Energy-Loss Spectroscopy (EELS) have allowed the acquisition of hyperspectral images (HSIs), (also known as spectral images – SI –), with typical sizes of several tens of thousands of pixels and around one thousand energy channels.

The fine structure of the characteristic edges provides access to the bonding environment and electronic structure of the elements constitutive of the sample. To interpret the data from a materials science point of view, the characteristic components and maps from the HSI need to be extracted. This extraction can be accomplished by processing each spectrum individually, for example, by subtracting the background and adding the characteristic signal corresponding to a given edge. However, looking at the HSI as a whole and taking a statistical view of the data is more efficient as well as more systematic and relevant regarding results[1].

One type of data processing application in STEM-EELS involves dimensionality reduction with Principal Components Analysis (PCA). The information contained in the HSI can be reduced to a few components, as spectral vectors lie very close to a low-dimensional subspace. One limitation of PCA involves the non-physical characteristics of the components extracted, which makes a physical interpretation tricky. Although expressed over the same spectral range, these components are not spectra, strictly speaking. Thus, more processes beyond PCA are necessary to provide a complete data processing result, i.e. a set of reference spectra and corresponding maps that can be used to support an interpretation.

An HSI is usually processed through the traditional background subtraction (BS) method and signal integration. The requirement separates a characteristic edge (for example, Co-$L_{2,3}$ edge) from its underlying background. The background is approximated to a power law energy dependence $AE^{-r}$, E being the energy loss and A, r, two parameters to be measured over a fitting region immediately preceding the edge. Once the background has been removed under the characteristic edge, the signal is integrated over the energy window of interest.

Despite its simplicity, this procedure has some drawbacks. For example, the fitting window and the integration window have to be carefully chosen, and these choices may introduce a user bias. Another issue is that the procedure cannot be applied when two edges overlap[3] or the same element is present with different electronic structures (valence, coordination) that we want to distinguish[2].

Consequently, it seems more efficient to fully exploit the low dimensionality structure of the HSI used in PCA and perform a form of data analysis that directly provides the desired result, as unmixing algorithms can extract significant components of the sample and compute associated maps[2,3]. The spectrum collected at an individual pixel is usually a mixture of the signatures of the different atoms interacting with the beam. Mixed pixels occur if the spatial resolution is low or if different compounds are present in the sample thickness intersected by the electron probe (e.g., particles in a matrix, diffusion at an interface, an atomic column with different elements), leading to an impure spectrum. Many techniques have been suggested to unmix the impure spectrum and recover the pure signals corresponding to the individual components of the sample.

A standard technique is linear spectral unmixing, which assumes that an individual spectrum is a linear combination of pure spectra[4]. In the case of EELS spectroscopy, a pure spectrum can correspond to one element or an element with a specific structural and electronic environment. For example, in[5], one seeks to separate the signal of Fe in a six-fold (octahedron) and Fe in a five-fold (distorted tetragonal pyramid) oxygen coordination. A pure spectrum can also contain two different elemental thresholds, as in[3]: one pure spectrum with both Ti-$L_{2,3}$ and O-K and another with Sn-M and O-K.

While the pixel size for an EELS SI is typically 0.05 nm for atomic resolution, at a completely different scale (about 1 m per pixel), remote sensing (use of satellite- or aircraft-based sensor technologies to detect and classify objects on Earth) produces HSIs with a data structure identical to STEM-EELS SI. Due to the importance of military, intelligence, commercial, economic, planning, and humanitarian applications, numerous frameworks have been developed to analyse vast quantities of data[4]. The STEM-EELS community can thus benefit from these results.

Many recent publications have discussed novel Deep Learning techniques[6-9] and applied them to processing remote sensing data. Applied to grayscale or colour images, Deep Learning methods use datasets that include thousands of images (70,000 for MNIST, 1,500,000 for ImageNet). In contrast, in the case of hyperspectral remote sensing images, access is only available for individual images. The subsequent dataset consists of a single HSI, where each pixel represents an item (or a group of pixels for methods that incorporate the spatial structure of the HSI). The training is then performed on the dataset defined by all the pixels of the HSI. Thus, there is no need to rely on an entire library of HSI.

Some interesting results have been obtained for denoising and classification in remote sensing[10]. In particular, some algorithms, called *autoencoders* (AE), are based on the principle of an encoding-decoding architecture. AEs represent a form of unsupervised learning with a loss function that compares the reconstructed spectrum to the original spectrum for each pixel. Moreover, with a specific AE architecture, it is possible to perform spectral unmixing, and several algorithms have been proposed[11-24]. EndNet[16] is based on a two-staged AE network with additional layers and a particular loss function. DAEN[17-19] is an AE consisting of two parts: a stacked AE for initialisation and a variational AE for unmixing.

The case of non-linearity can be addressed by adding a non-linear component to the decoder[11, 13, 14]. In work[13], these networks are improved by incorporating the spatial structure of the data using a 3D-Convolutional Neural Network (CNN). New works have combined this spectral-spatial information with architectures designed to cope with the endmember variability[25, 26]. An adaptation of the architecture used in[12] is presented in[15]. Recently, a transformer network has been combined recently with a convolutional AE to capture the interaction between image patches[27].

An occurrence of unmixing AE appeared in[21] and was developed in[22]. The work[23], using an architecture inspired by multitask learning, operates on image patches instead of single pixels to utilise the spatial structure. CNN is used in[24] to capture the spatial correlations existing in HSIs.

This article does not include a complete list of references, as the number of studies devoted to AEs applied to SU has increased rapidly in recent years. Only some of the previously described codes are publicly available to perform unmixing with AEs, although, a series of codes have recently been made available to the community[28].

To evaluate the performance of these methods as applied to STEM-EELS HSIs, state-of-the-art and often quoted models that are among the publicly available ones were selected, including uDAS[20], deep AE unmixing (DAEU)[22], multitask AE (MTAEU)[23] and CNN AE (CNNAEU)[24]. These algorithms are presented in section 3. The performances of these algorithms are compared to those of conventional unmixing algorithms currently used in the STEM-EELS community, such as Independent Component Analysis (ICA), Non-Negative Matrix Factorization (NMF) (as implemented in the popular toolbox Hyperspy[29]), Vertex Component Analysis (VCA)[30] that appears at the moment as the most versatile algorithm to perform spectral unmixing, and BLU[2, 31], which is a Bayesian algorithm that estimates the endmembers and the abundances jointly in a single step.

Deep-learning algorithms need to be verified before they can replace the traditional SU techniques. They nevertheless hold the potential to improve the quality of the results, as well as the execution time. A neural network can be long to train but, its inference is high-speed if applied to different data sets, such as a series of HSIs acquired on the same sample, or similar samples during an acquisition session on a given microscope. It is essential to compare the performance of the different algorithms quantitatively.

Synthetic datasets were generated using the method described in[32] to provide this quantitative assessment. These algorithms were then applied to an experimental dataset. As no ground truth (GT) is available for this dataset, only a qualitative evaluation was performed using the chemical maps obtained by the usual BS method.

The remainder of the paper is organised as follows. Section 2 describes the synthetic datasets and metrics used to quantitatively evaluate the different unmixing algorithms. Section 3 briefly presents the different algorithms used and the results obtained for the synthetic datasets. Section 4 applies the same algorithms to real SI datasets. Finally, section 5 is the conclusion.

## 2. Synthetic datasets and metrics

The performances of different state-of-the-art unmixing methods were compared with those of Deep Learning based methods using synthetic data. The synthetic data was generated with the linear mixture model. Two sets of endmembers were used, one with three endmembers and the other with four endmembers. Endmembers were extracted from the experimental dataset of section 4 for the three components HSI (Figure 1) and obtained from data described in[2,33] for the four components HSI (Figure 2).

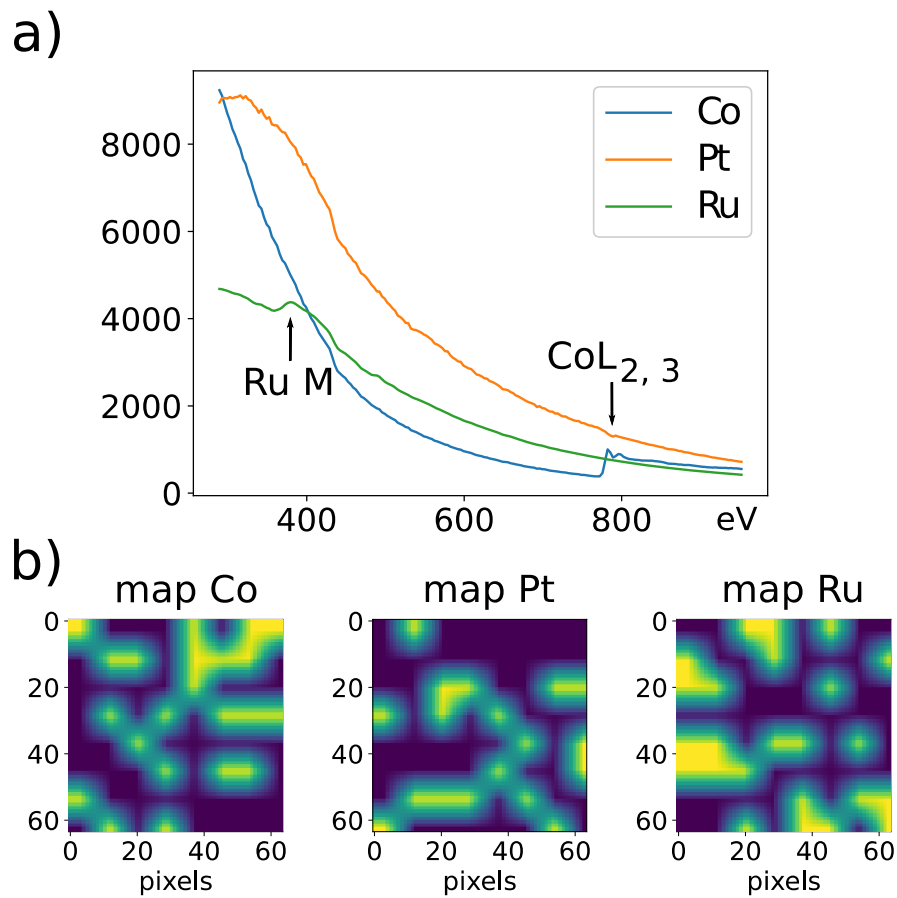

Figure 1. a) The three components set used to create a synthetic spectrum image - they have been extracted from an experimental dataset, b) c) d) 64 x 64 maps obtained with the chequerboard model (8x8 blocks); each of these maps is associated with one component.

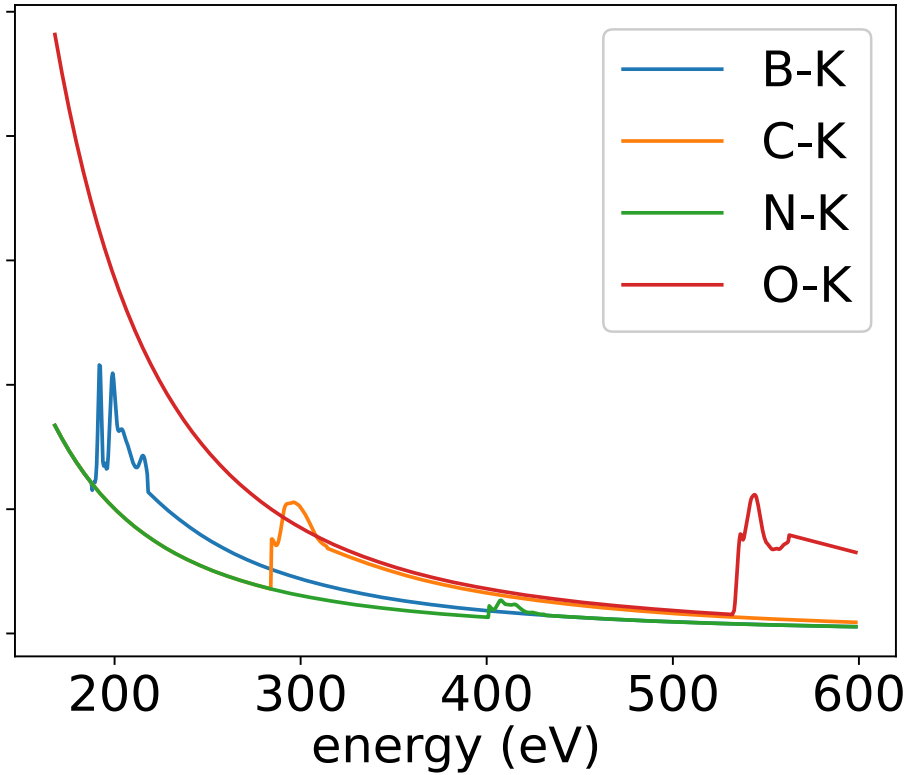

Figure 2. The four components set used to create a synthetic spectrum image. They have been obtained by fitting spectra extracted from the dataset used in[2, 29]

The three components set has 200 energy channels, and the four components set has 903 energy channels. For the component referenced as "Pt", no Pt edge is present in this energy range, as the Pt-M edge energy is 2122 eV. The Pt component here is only the background in the part of the sample containing Pt.

For the maps, images of 64x64 pixels were divided into units of 8x8 small blocks, as described in[32]. Each block was randomly assigned to one of the endmembers. A $k \times k$ averaging filter then degraded the resulting image to create mixed pixels, with $k=9$, with abundances respecting the sum-to-one rule (Figure 1).

The maps were combined with the components following the LMM to create a synthetic SI. Finally, Poissonian noise was applied to the SI using the in-built Hyperspy method. Poissonian noise was chosen in preference to Gaussian noise because the increasingly frequent use of direct detection cameras provides data degraded by this type of noise, as is the case for experimental data from section 4.

For quantitative comparison unmixing results from the synthetic data were evaluated using two metrics, Spectral Angle Distance (SAD) and Normalised Mean Squared Error (NMSE). SAD, a widely used metric in HSI analysis, was used to measure the similarity between the extracted endmember and the true endmember (GT).

SAD is expressed as:

$$SAD(a, \hat{a}) = \cos^{-1}\left(\frac{a^T \hat{a}}{\|a\|_2 \|\hat{a}\|_2}\right)$$

with $a$ the spectral vector corresponding to the true endmember (GT), $\hat{a}$ its estimation by the Neural Network, $a^T$ is $a$ transposed, and $\|.\|_2$ is the $\ell_2$ norm.

NMSE measured the performance of abundance estimation defined as:

$$NMSE(Z, \hat{Z}) = \frac{\|Z - \hat{Z}\|_F^2}{\|Z\|_F^2}$$

with $Z$ the true abundance of a given endmember (it is a matrix) and $\hat{Z}$ its estimation by the Neural Network, $\|.\|_F$ being the Frobenius norm defined as:

$$\|Z\|_F = \sqrt{\sum_{\substack{1 \leq i \leq n \\ 1 \leq j \leq m}} |Z_{ij}|^2}$$

with *n, m* being the size of the spatial dimensions of the data cube and $|Z_{i,j}|$ the absolute value of matrix element $Z_{i,j}$. SAD and NMSE are frequently used in the literature on hyperspectral unmixing[13, 26, 28] and sufficiently characterise our results.

For each set of endmembers 20 different series of three (respectively four) chequerboard images were generated using the method described previously. Thus, for each algorithm tested, a mean and a standard deviation were obtained for each evaluation by a given metric.

## 3. Unmixing algorithms

Before proceeding to unmixing, the number of endmembers has to be estimated, as this is a parameter used by the unmixing AE. Although various algorithms already exist for this estimation[34], the process remains a delicate task. The estimation is all the more difficult as the definition of endmembers is subjective and depends on the degree of precision to be attained. For example, if the Co in the sample had different degrees of oxidation, it is not clear if it would be necessary to distinguish metallic Co from $Co^{2+}$ and $Co^{3+}$. The number of endmembers also depends on the level of information sought and the experimental conditions of data acquisition (energy resolution for EELS spectroscopy). Estimating the number of endmembers is particularly difficult in remote sensing, where the sensor can uncover unknown target sources that cannot be identified a priori. However, this estimation process is different in microscopy, where the global composition of the sample, which is a manufactured material, is well known. On the other hand, including more endmembers will not immediately improve the quality of unmixing results. Thus in ref[28], increasing the number of endmembers for the Urban image unmixing degrades the results for most tested AEs.

In this study, the number of endmembers for the synthetic images was the same as the number used to generate the data. As the sample composition was known for the experimental data, this physical information was part of the unmixing process. Thus each element present in the sample, i.e., Ru, Co and Pt, corresponded to one spectral signature and then to one endmember. This choice of the number of endmembers implied that any spectral variation occurring in the data was assimilated to noise.

The methods compared in this work are listed in Table 1.

| Name | Architecture | Implementation |
| --- | --- | --- |
| NMF | Non-negative Matrix Factorisation | Hyperspy[29] |
| ICA | Independent Component Analysis | Hyperspy[29] |
| VCA[30] | Geometrical based approach | code available on the web |
| BLU[31] | Statistical Bayesian | code available on the web |
| DAEU[22] | Deep AE | code available on the web |
| MTAEU[23] | Deep encoder utilising MTL | code available on the web |
| CNNAEU[24] | fully CNN encoder and decoder | code available on the web |
| uDAS[20] | untied denoising AE with sparsity | code available on the web |

Table 1. Methods used in this work (adapted from[28])

Although many spectral unmixing algorithms exist, unmixing is not frequently used in the STEM-EELS domain. The algorithms most used are NMF and ICA - as implemented in the Toolbox Hyperspy - and VCA.

Hyperspy uses the Scikit-learn version of the NMF algorithm, `sklearn.decomposition.NMF`[36]. NMF decomposes the data matrix **X** by resolving the optimisation problem:

$$\min(L(\mathbf{X}, \mathbf{WH})) \; with \; \mathbf{W}, \mathbf{H} > 0$$

**X** being the data, L a loss function measuring the similarity between **X** and the decomposition, and **W** and **H** the matrices resulting from the decomposition. A regularisation term can be added to the loss function. Furthermore, the initialisation mode can be changed, as well as the distance. The default values provided in Hyperspy were kept (Frobenius norm, no regularisation, initialisation by Non-negative Double Singular Value Decomposition).

ICA decomposes **X**, assuming that the components are statistically independent[37]. Independency is a strong assumption whose relevance has been questioned in the remote sensing community[30, 37]. It is a widely used method in STEM-EELS, probably due to its implementation from the first versions of Hyperspy. Hyperspy uses, by default, the Scikit-learn implementation, `sklearn.decomposition.FastICA`. This implementation is based on[38]. This default version was used in the present work.

VCA is one of the most advanced convex geometry-based endmember detection methods. It is based on successive projections on hyperplanes[30]. This algorithm assumes the presence of at least one pure pixel for each component in the data. If there is no pure pixel, it uses the highest quality pixel that is available. Although the pure pixel condition is not always verified in STEM-EELS HSIs, this algorithm is fast and computationally relatively light. VCA has been implemented in Python (`https://github.com/Laadr/VCA`) and Matlab

(`http://www.lx.it.pt/~bioucas/code.htm`). It is commonly used in the remote sensing community and is often used as a reference to evaluate new unmixing algorithms. It has already been used to unmix STEM-EELS SI[2, 39, 40].

BLU is a fully Bayesian algorithm which uses a Gibbs sampler algorithm to solve the unmixing problem without requiring the presence of pure pixels in SI[31]. Its performance for EELS HSIs has been evaluated in[2].

AE*s* are an unsupervised learning technique using neural networks to learn a latent space representation of the input. The AEs have a small number of layers, and they belong to the domain of representation learning rather than traditional Deep Learning. The part of the network that compresses the input into the latent representation is called the encoder. The part that reconstructs the input from the latent representation is the decoder. The AE architecture imposes a bottleneck that forces a compressed input representation (Figure 3).

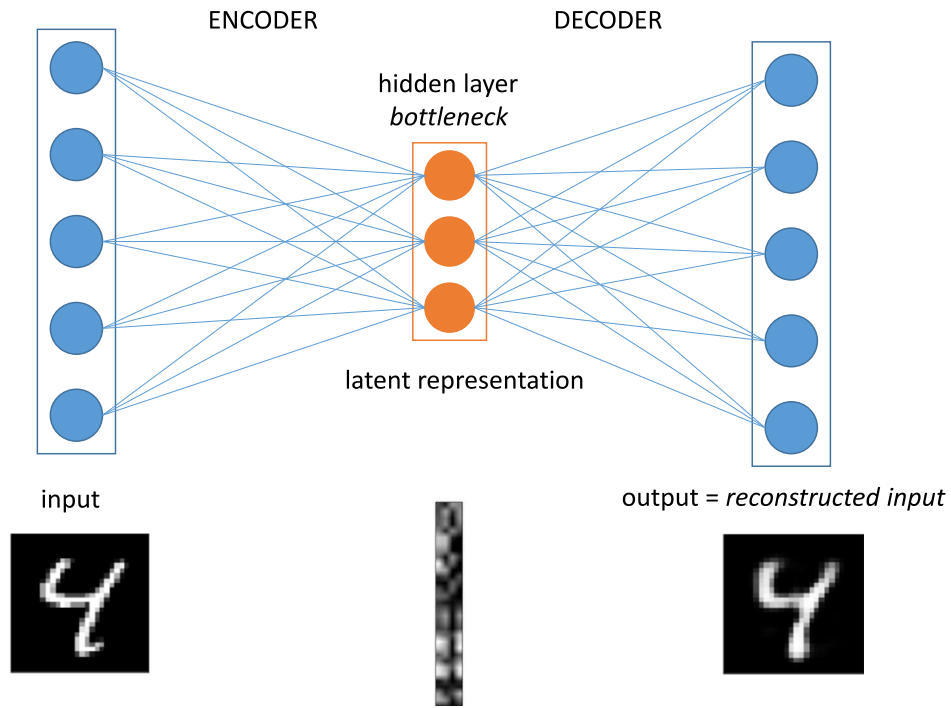

Figure 3. Basic principle of an AE. The hidden layer is a bottleneck that forces a compressed input representation. The input example is a 2D image from the MNIST dataset. The 2D 28x28 pixels image is processed as a 784 vector. In our case, the input is a spectrum.

The AE is trained by using the input as target data, meaning that the AE learns to reconstruct the original input. The decoder part of the AE aims to reconstruct the input from the latent space representation. By limiting the decoder to one layer, it has been shown that the activations of the last layer of the encoder correspond to the abundances and the weights of the decoder to the endmembers[22]. The encoder converts the input spectra to the corresponding abundance vectors, i.e., the output of the hidden layer. The decoder reconstructs the input from the compressed representation with the weights in the last linear layer interpreted as the endmember matrix[28]. The action of the last layer of the decoder can be written as: $\hat{x}_p = W^{(L)} a^{(L-1)}$ where $\hat{x}_p$ is the output of the network (reconstructed spectrum), i.e., an estimation of the input $x_p$, $a^{(L-1)}$ are the activations of the previous layer, $W^{(L)}$ are the weights of the output layer, $L$ being the total number of layers, B the number of bands and $R$ the number of endmembers.

$a^{(L-1)}$ is of dimension $Rx1$, and $W^{(L)}$ is a $BxR$ matrix, which has to be interpreted as abundances and endmembers for a given input. The weights are fixed once the network is trained, and the endmembers are determined for the whole dataset. The activations are dependent on the input (pixel) analyzed.

According to this principle, the decoder must be a single layer, and this simple structure might affect the performance of the AE; however, the experiments show that this AE performs well in unmixing the remote sensing data. Although several articles have proposed neural networks to achieve unmixing, the corresponding code is only sometimes published in parallel. This lack of information is detrimental, as not only is re-implementing the code time-consuming, but many implementation details, such as utility layers and hyperparameters values, are not specified in publications on the subject, while modifying these features can significantly alter the results. Recent efforts have started to mitigate this issue, and several codes are available on the web, e.g. uDAS on GitHub (`https://github.com/aicip/uDAS`). Although linked to the field of Deep learning by its keywords, uDAS is a shallow AE with only one encoding layer. Its architecture makes it somewhat close to a conventional optimisation method for an inverse problem, with an alternating optimisation of the encoder and the decoder[20]. Implementations of DAEU (see Table 2), MTAEU and CNNAEU have been

recently made available

(https://github.com/burknipalsson/hu_autoencoders) with the corresponding publication[28].

| layer# | layer type | activation | units# |
|---|---|---|---|
| 1 | Input | - | $B$ |
| 2 | Dense | LReLU | $9R$ |
| 3 | Dense | LReLU | $6R$ |
| 4 | Dense | LReLU | $3R$ |
| 5 | Dense | LReLU | $R$ |
| 6 | Batch Normalization | Utility | $R$ |
| 7 | Dynamical Soft Thresholding | LReLU | $R$ |
| 8 | ASC enforcing | Utility | $R$ |
| 9 | Gaussian Dropout | Utility | $R$ |

Table 2. Detail of layers of the encoder in DAEU network[22]. $B$ is the number of channels of the input (spectrum). $R$ is the number of units of the latent, hidden layer, i.e., the number of components to unmix. The utility layer performs an operation not specific to a neural network; in particular, the utility layer does not change the number of units.

The models for all the AEs evaluated in this work were those available online without changing the architecture or the hyperparameters.

**Unmixing of synthetic data**

The results of the different unmixing algorithms are shown in Figure 4 and Figure 5.

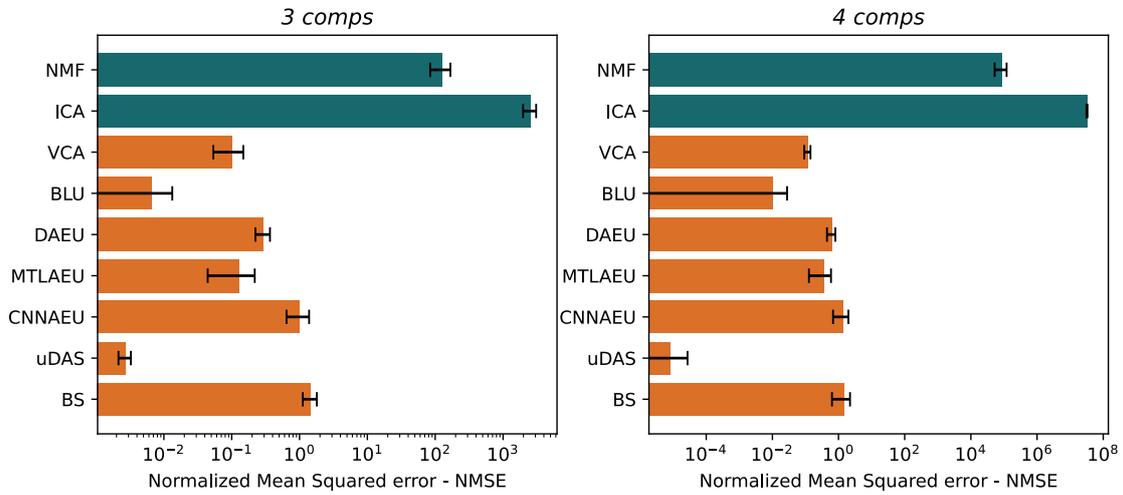

Figure 4. Performance of different algorithms for abundances maps estimation with a log scale. 'BS' is the background subtraction method.

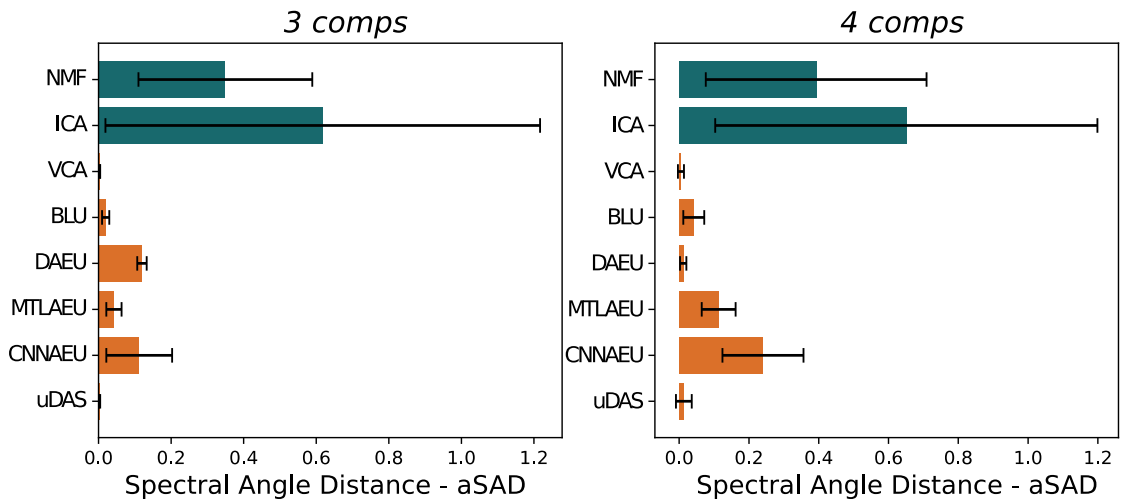

Figure 5. Performance of different algorithms for endmembers extraction.

uDAS is the algorithm that gives the best results for both sets of synthetic data, three and four components. In contrast, ICA and NMFgive the worst results. The others (VCA, BLU, DAEU, MTAEU, CNNAEU) produce intermediate-quality results. uDAS has the particularity of including denoising and regularisation constraints ($\ell_{2,1}$ on endmembers), which may explain the good results obtained. It should be noted that the chequerboard structure of synthetic images creates a non-zero proportion of pure pixels, which favours the use of VCA and uDAS (as it is initialised with VCA).

The results are qualitatively the same for ranking the algorithms in terms of performance for the two types of synthetic data: three components with 200 energy channels (Figure 1) and four components with 903 energy channels (Figure 2). Moreover, the spectra chosen to build the data are significantly different, with M and L edges (Figure 1) versus K edges (Figure 2), which does not influence the results significantly.

The SAD metric is scale-invariant and solely takes into account the shape of an extracted component. Its absolute amplitude can be very different from the reference endmember without affecting the result. In contrast, the NMSE metric will calculate a significant error for a calculated abundance map that does not respect the sum-to-one rule. NMF and ICA methods do not use this constraint, so they get a high error.

No significant improvement is observed in MTAEU and CNNAEU compared to DAEU, despite their higher complexity in accounting for the HSI spatial correlations between pixels. These methods utilize this spatial information by treating the input as a patch (a patch refers to a small block of pixels, typically a 3x3 square with 9 pixels).

Thus, while patch-based methods usually outperform conventional methods in image analysis, in this case spatial structure does not have an impact on the quality of unmixing for STEM-EELS HSIs.

The higher number of hyperparameters related to a complex architecture may require adjustment to the characteristics of the STEM-EELS HSIs, i.e., more energy channels and fewer pixels.

The computation times required by each algorithm are reported in Table 3 (3 GHz Intel® Core™i7-1185G7 - except for CNNAEU, which has been trained on a computer with a GPU NVIDIA Quadro RTX4000 8Go (7.5 Cuda score)). The complexity of VCA, ICA, and NMF is lower than those of DAEU, BLU and uDAS.

| unmixing method | 3 components-200 channels SI | 4 components-900 channels SI |
| --- | --- | --- |
| NMF | 2.2 | 5.2 |
| ICA | 0.1 | 0.1 |
| VCA-FCLS | 0.1 | 0.1 |
| BLU | 659 | 2055 |
| DAEU | 84 | 122 |
| MTAEU | 192 | 612 |
| CNNAEU | 491 | 1726 |
| uDAS | 72 | 1021 |

Table 3. Execution time (expressed in s) for different algorithms on synthetic data. For the AE, the time is the training time.

As the acquisition of the data is relatively fast, around 10 minutes for core loss data, even less with the new generation of direct detection detectors, the microscope user might want to process the data quickly, whether done after or online during the experiment on the microscope. During the acquisition time, using basic neural networks could allow the training to be carried out with a first data cube (or a previous one in the case of a series of experiments). Then one could apply the trained network to the following acquisitions, reducing the execution time to 0.1 seconds. The experimental conditions of STEM-EELS are thus particularly well adapted to using a neural network because these networks allow exploiting several HSIs acquired under the same conditions. This situation differs from the case of HSIs acquired in remote sensing, where the cases presented in the literature correspond to the exploitation of a single HSI. Even if the performance of AEs for spectral unmixing is currently limited, their use remains interesting in STEM-EELS because of their speed in inference.

## 4. Experimental dataset

The different algorithms were applied to a HSI acquired on a Pt/Co/Ru/Pt multilayer[41]. These heterostructures were investigated regarding their magnetic properties, i.e. Dzyaloshinskii-Moriya interaction, at metallic interfaces. The nominal stacking corresponds to: Si/SiO$_2$/Ta(10 nm)/Pt(8 nm)/Co(1.7 nm)/Ru(0.5 nm)/Pt(3 nm). In these samples, the Ru layer and its top and bottom interfaces, i.e. Pt/Ru and Ru/Co, respectively, can have an impact on the local magnetic properties of the stacking. Therefore, characterising this layer and the corresponding interfaces is essential. In particular, the Ru can diffuse into the Co layer, and it is thus necessary to establish a profile for Ru.

The data were acquired on a USTEM Nion microscope operated at 100 kV using a Medipix3 detector (Merlin EM Quantum Detector) with a 50 ms dwell time. The initial HSI is 60x120 pixels x 200 energy channels (SI2 in Fig. 6). A pixel represents 0.12 x 0.12 nm and an energy channel 3.33 eV. The image SI1 is cropped from SI2 and represents 60 x 75 pixels. Data were corrected for gain before any advanced post-treatment.

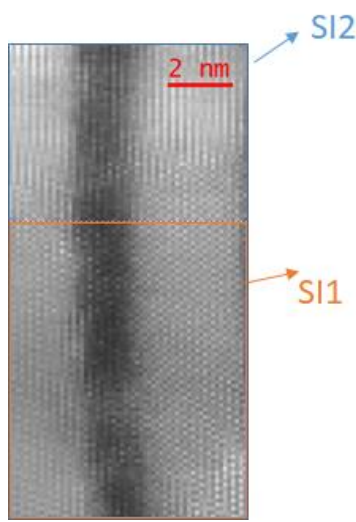

Figure 6. HAADF image corresponding to the HSI. A sub-area representing 60x75 pixels - SI1 - has been extracted from the 60x120 pixels image - SI2 -.

As the Ru is a delayed M-edge (279 eV for $M_{4,5}$), it was challenging to determine the maps using the BS and characteristic signal integration method used in EELS. Therefore, using an unmixing method for this type of data was interesting. Nevertheless, the $M_{2,3}$ edge of Ru was used to obtain intensity maps to compare with the maps obtained by unmixing. (Figure 7), as the $M_{2,3}$ edge is detectable with the direct detection camera.

The profiles in Figure 7 are obtained by summing three lines of pixels corresponding to 3.6 nm width at the top of the HSI (SI1). There is an artefact with a small non-zero intensity outside the Ru layer, depending on the pre-edge energy window selection.

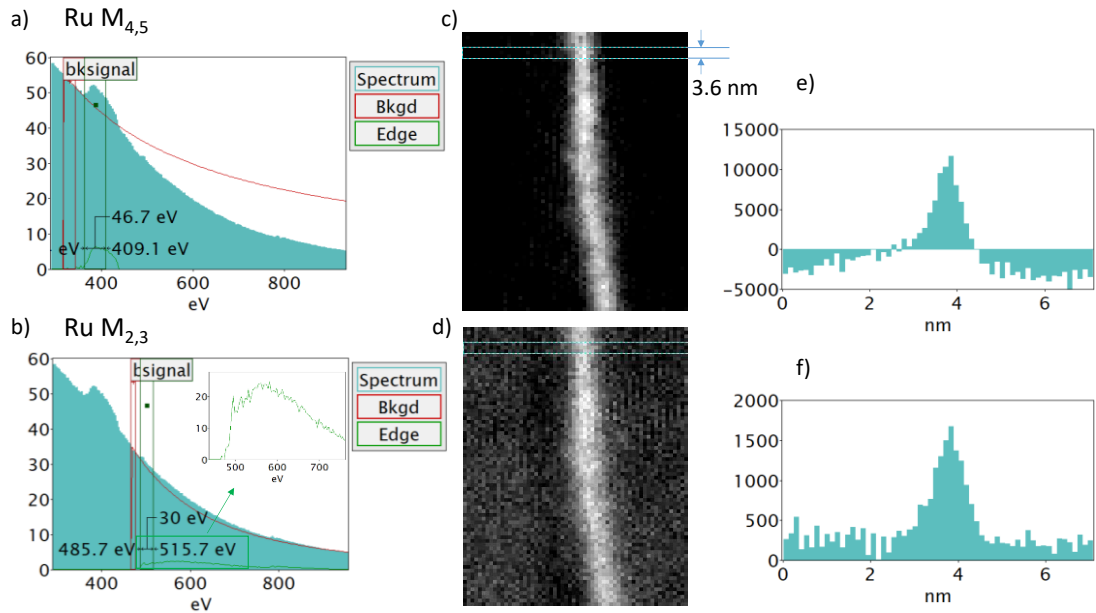

Figure 7. Using the $M_{4,5}$ edge -a) - to map the Ru - c) - produces negative values for intensities, as can be seen in the resulting profile - e) -. The $M_{2,3}$ edge - b) - gives better results - d) -. on the profile - f) – there is an artefact with a small non-zero intensity outside the Ru layer.

The Pt-M edge is in a high-energy range (2122 eV), so this edge was not used in the unmixing process.

As it is a real sample, there is no available GT, so it was impossible to compute metrics for experimental data; then, the evaluation was qualitative. However, the profiles obtained by unmixing were compared with those obtained by the BS method. The following results are obtained with the processing of data of SI1, which is a restricted area of 60x75 pixels. The extracted components are presented in Figure 8.

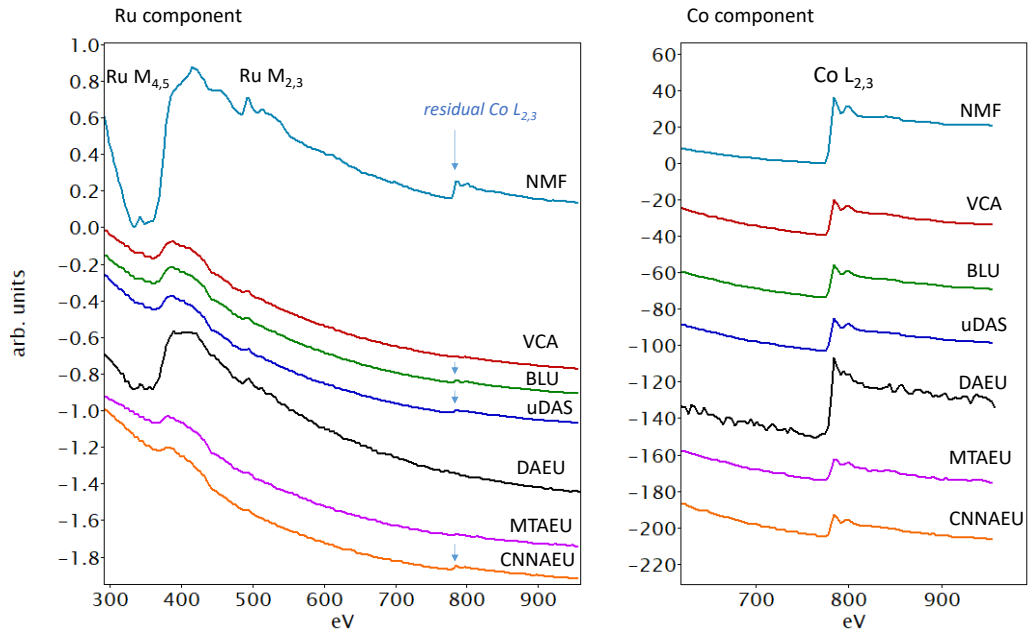

Figure 8. Components obtained by the different unmixing algorithms tested on SI1. Each component corresponds to a unique edge, Ru-M and Co-$L_{2,3}$.

The Co component is well-extracted in all cases. The components obtained by VCA, BLU and uDAS are very similar, probably because uDAS and BLU are initialised by VCA.

Ru component extraction is more challenging, and there is still a Co signal in all components except in the component extracted by DAEU. VCA relies on the pure pixel hypothesis, and there is probably no pure pixel corresponding to Ru. A remarkable result is that the DAEU neural network not only manages to remove the Co but obtains a component close to the reference edge obtained in[42]. Despite a very low dispersion (about 3.3 eV/channel), some fine structure is present on the Ru-M edge. Although they have much more complex architectures, MTAEU and CNNAEU do not manage to extract the Ru component more satisfactorily than DAEU in the case of SI1. The resulting maps are presented in Figure 9.

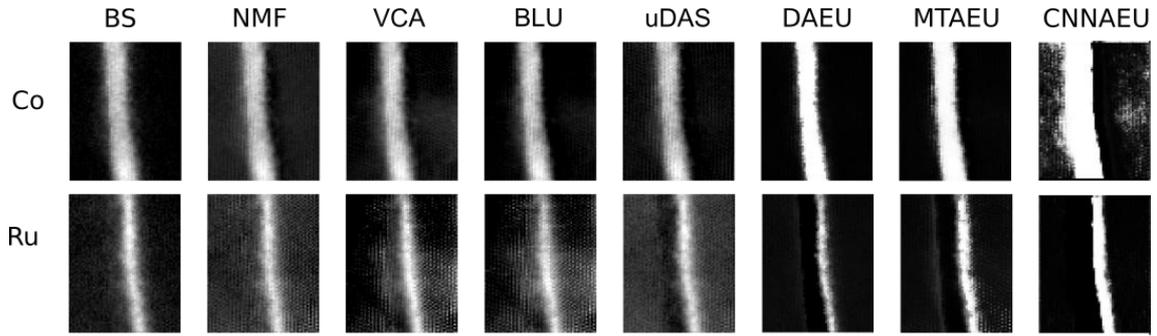

Figure 9. Maps obtained by the different unmixing algorithms tested on SI1. As each component corresponds to a unique edge, they are elementary maps.

The maps obtained with NMF are visually the closest to those obtained by BS. The maps obtained by VCA, BLU and uDAS are satisfactory. DAEU, MTAEU and CNNAEU give very contrasting maps with a steep interface between the Co and Ru layers, which does not correspond to the physical reality. For DAEU and MTAEU, the abundances are close to either 0 or 1; this problem has been reported in the literature for this class of AEs[23]. Despite their complexity, MTAEU and CNNAEU do not perform better than DAEU on the experimental data. It might be necessary to adjust some hyperparameters, which is a complex task, as the networks are trained on the reconstruction quality rather than on the quality of the unmixing.

To obtain a more precise estimation of the quality of the unmixing, the profiles obtained by the different unmixing methods are presented in Figure 10. The Ru profile presents an asymmetry with diffusion in the Co layer.

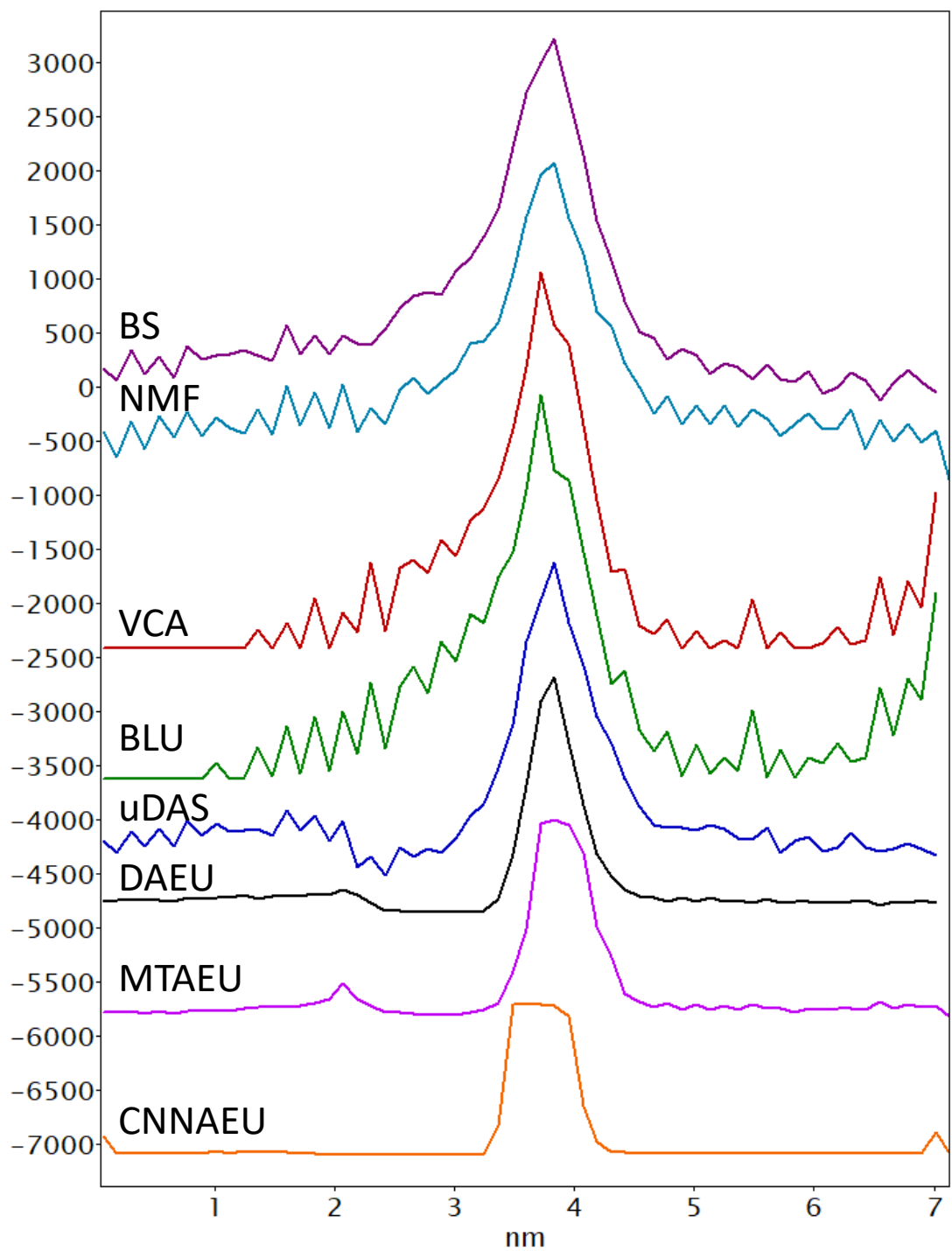

Figure 10: Profiles through Ru maps presented in Figure 9. Intensities values (arbitrary units) have been re-scaled to match the 'BS' profile.

If the unmixing is correctly performed, the shape of the profile should be close to those obtained by the BS method. The comparison of abundance profiles is a criterion that is not, to our knowledge, used in remote sensing, where one relies solely on metrics and the visual comparison of maps to evaluate the performance of different methods. As was the case for the maps, the profile obtained by NMF corresponds to the profile obtained by BS and appears close to physical reality. The profiles obtained by VCA and BLU are also satisfying. They are close to zero, far from the layer on the left part of the layer, while it was not possible to eliminate the signal by subtracting the background in front of the threshold. However, an anomaly is observed on the right side of the profile. Some degree of spectral variability (caused, for example, by variations in thickness) could explain why the algorithms have difficulty representing the data set with only three components.

uDAS somehow reproduces the asymmetry of the profile but shows a non-zero intensity away from the Ru layer. DAEU, MTAEU and CNNAEU fail to reproduce the asymmetry of the profile; thus, the weight of the Ru component falls to 0 in the region where it is mixed with Co. Therefore, the improved extraction of the Ru component by DAEU did not result in a satisfactory map.

In the case where the data present spectral variability, i.e., if slightly different spectra represent the same component, VCA (and other unmixing algorithms) performance is affected[42]. This problem is illustrated in Figure 11, where the unmixing results are presented for SI2, i.e. for an area larger than SI1 (Figure 6). SI2 shows changes such as variations in thickness, some degree of damage due to the preparation of the thin slide, or inhomogeneities of the sample due to the deposition process of the heterostructure.

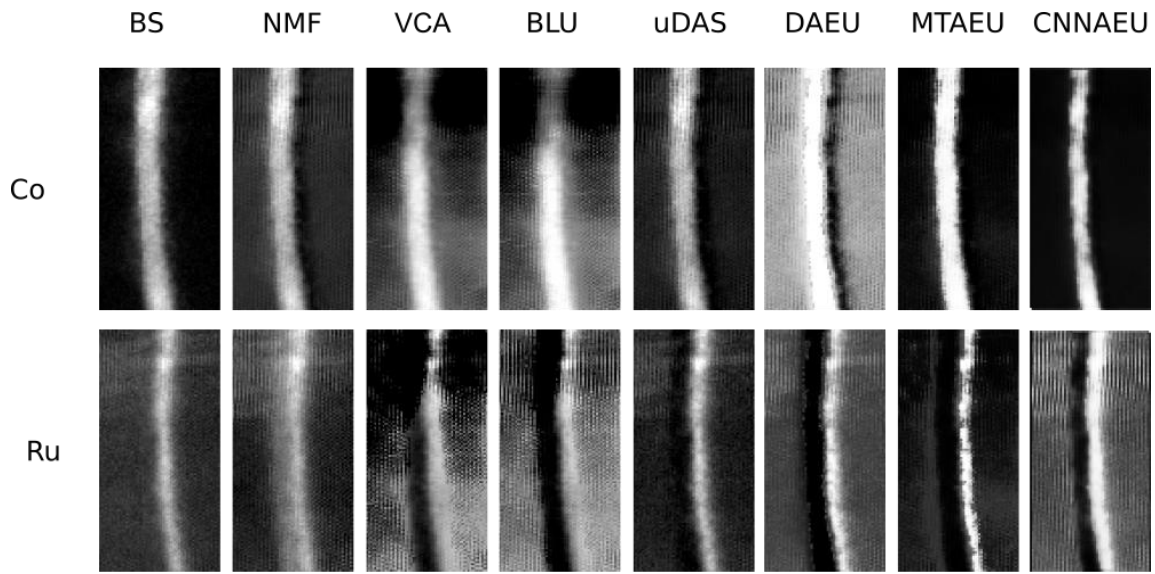

Figure 11. Maps obtained by the different unmixing algorithms tested on SI2 (SI1 in Figure 9 was cropped from SI2). Variations of spectral signatures make the unmixing process more challenging.

In this more complex case, the methods that performed well in the more straightforward case of SI1, i.e. NMF, VCA, BLU and DAEU, perform less well. The more complex methods, such as uDAS, MTAEU and CNNAEU, are then of interest.

Note that the VCA results could be improved by using a larger number of components for unmixing (typically six in this case) to account for spectral variability. However spectral variability makes interpreting the unmixing results in terms of elementary proportions more difficult.

The results obtained on the synthetic data are not fully applicable to the experimental data. Several limitations of synthetic SI can be identified. Firstly, these images are constructed very simply; spectral variability is not considered. Moreover, the synthetic chequerboard maps contain a certain proportion of pure pixels for each component. In contrast, in the experimental situation, one of the components (Ru) corresponds to almost no pure pixels. To elaborate synthetic data closer to reality would be necessary to obtain a relevant evaluation of the efficiency of the unmixing algorithms.

## 5. Conclusions

This work demonstrates that AEs give interesting results for spectral unmixing. In particular, suitable extraction of the Ru component can be obtained despite the absence of pure pixels for this element in the experimental data. Moreover, the organisation of the STEM-EELS experiments makes them well adapted to Deep Learning: the network is trained on the first set of data (first acquired HSI) and then the weights are applied to the data acquired subsequently while benefiting from a swift execution time. This procedure can also apply if a series of very similar samples is studied (for example, in the case of Pt/Co/Ru sample by varying the thickness of the Ru layer).

A simple neural network such as DAEU performs well on a homogeneous image such as SI1; the results are degraded on a larger area as SI2, which shows spectral variability. More complex neural networks such as CNNAEU and MTAEU, which are efficient according to the literature in remote sensing[28], should be able to handle STEM-EELS data. However, they output worse results on our STEM-EELS data. One hypothesis is that this failure is due to the specific shape of the EELS spectra with a strong signal represented by the continuous background and relatively weak superimposed specific signals. Adapting either the hyperparameters (batch size, number of hidden units...) or the architecture would probably be necessary.

On the other hand, the results obtained on the experimental data are not as good as expected from the first tests on the synthetic data. The model used to create synthetic

HSI is too simple, and introducing a degree of spectral variability would be helpful, for example by using a variational AE (VAE)[28] which encodes the input as a distribution.

Researchers continue to make progress on hyperspectral unmixing by Neural Networks. More efficient networks will allow us to consider the spectral variability, and such progress can be achieved through cooperative work in the field, in particular by allowing open access to the codes used[10].

*Acknowledgements*

We thank André Thiaville (LPS, Orsay), William Legrand, Nicolas Reyren and Vincent Cros (UMP CNRS/Thales, Palaiseau) for the Pt/Co/Ru data. We are grateful to Calvin Peck from Academic Writing Center (U. Paris Saclay) for his patience in editing the manuscript.

This project has been funded in part by the National Agency for Research under the program of future investment TEMPOS-CHROMATEM (reference no. ANR-10-EQPX-50) and by the European Union's Horizon 2020 research and innovation program under grant agreement No. 823717 (ESTEEM3)

**Figures Captions**

Figure 1. a) The 3 components set used to create a synthetic spectrum image - they have been extracted from an experimental dataset, b) c) d) 64 x 64 maps obtained with the chequerboard model (8x8 blocks); each of these maps is associated with one component.

Figure 2. The four components set used to create a synthetic spectrum image. They have been obtained by fitting spectra extracted from the dataset used in[2, 29]

Figure 3. Basic principle of an autoencoder. The hidden layer is a bottleneck that forces a compressed input representation. The input example is a 2D image from the MNIST dataset. The 2D 28x28 pixels image is processed as a 784 vector. In our case, the input is a spectrum.

Figure 4. Figure 4. Performance of different algorithms for abundances maps estimation with a log scale. 'BS' is the background subtraction method.

Figure 5. Performance of different algorithms for endmembers extraction.

Figure 6. HAADF image corresponding to the HSI. A sub-area representing 60x75 pixels - SI1 - has been extracted from the 60x120 pixels image - SI2 -.

Figure 7. Using the $M_{4,5}$ edge -a) - to map the Ru - c) - produces negative values for intensities, as can be seen in the resulting profile - e) -. The $M_{2,3}$ edge - b) - gives better results - d) -. on the profile - f) – there is an artefact with a small non-zero intensity outside the Ru layer.

Figure 8. Components obtained by the different unmixing algorithms tested on SI1. Each component corresponds to a unique edge, Ru-M and Co-$L_{2,3}$.

Figure 9. Maps obtained by the different unmixing algorithms tested on SI1. As each component corresponds to a unique edge, they are elementary maps.

Figure 10: Profiles through Ru maps presented in Figure 9. Intensities values (arbitrary units) have been re-scaled to match the 'BS' profile.

Figure 11. Maps obtained by the different unmixing algorithms tested on SI2 (SI1 in Figure 9 was cropped from SI2). Variations of spectral signatures make the unmixing process more challenging.

**Tables Captions**

Table 1. Methods used in this work (adapted from[28])

Table 2. Detail of layers of the encoder in DAEU network[22]. $B$ is the number of channels of the input (spectrum). $R$ is the number of units of the latent, hidden layer, i.e., the number of components to unmix. The utility layer performs an operation not specific to a neural network; in particular, the utility layer does not change the number of units.

Table 3. Execution time (expressed in s) for different algorithms on synthetic data. For the AE, the time is the training time.